\documentclass[pra,reprint,floatfix,showpacs]{revtex4-1}

\usepackage{amsmath,amsfonts,amssymb}
\usepackage{graphicx}

\def\beq{\begin{equation}}
\def\eeq{\end{equation}}
\def\beqn{\begin{eqnarray}}
\def\eeqn{\end{eqnarray}}

\def\br {{\bf r}}

\begin{document}

\title{Spatially partitioned many-body vortices}

\author{Shachar Klaiman$^{1}$ and Ofir E. Alon$^2$}

\affiliation{$^1$ Theoretische Chemie, Physikalisch--Chemisches Institut, Universit\"at Heidelberg, 
Im Neuenheimer Feld 229, D-69120 Heidelberg, Germany}
\affiliation{$^2$ Department of Physics, University of Haifa at Oranim, Tivon 36006, Israel}

\date{\today}

\begin{abstract}
A vortex in Bose-Einstein condensates is a localized object 
which looks much like a tiny tornado storm.
It is well described by mean-field theory.
In the present work we go beyond the current paradigm and
introduce many-body vortices.
These are made of {\it spatially-partitioned} clouds,
carry definite total angular momentum, 
and are fragmented rather than condensed objects
which can only be described beyond mean-field theory.
A phase diagram based on a mean-field model assists 
in predicting the parameters where many-body vortices occur. 
Implications are briefly discussed.
\end{abstract}


\maketitle 

A vortex is a fundamental object in many branches of physics and engineering.
Vortices have been created in experiments with Bose-Einstein 
condensates (BECs) made of ultracold quantum gases \cite{ex1,ex2,ex3,ex4}.
There are ample theoretical works discussing vortices, variants thereof, and their properties
in BECs, see, e.g., 
\cite{vx1,vx2,vx3,vx4,vx5,vx6,vx7,vx8,vx9,vx10,vx11,vx12,vx13,vx14,vx15,vx16,vx17,vx18}.
For a comprehensive review see \cite{Fetter_Rev} and references therein.

BECs are made of many identical particles.
As such, they are subject to the Schr\"odinger equation
and are described by a \textit{many-body} wavefunction.
In BECs, 
a vortex is commonly described by 
a complex order parameter (single-particle function) which is the solution of the non-linear Schr\"odinger equation, 
or, as it is explicitly known to in the field,
Gross-Pitaevskii equation \cite{Fetter_Rev,Book_Pitaevskii,Book_Leggett,Book_Pethick}.
This implies that a vortex is a quantum object for which 
the many-body wavefunction is well approximated by the product
\beq\label{GP_Wf}
\phi(\br) = e^{il\varphi} f(r) \ \ \longrightarrow \ \
\Phi(\br_1,\ldots,\br_N) = \Pi_{j=1}^N \phi(\br_j).
\eeq
In (\ref{GP_Wf}), $l$ is an integer which equals
to the angular momentum carried by each particle
and $N$ is the number of particles.
For simplicity, the coordinate $\br=(r,\varphi)$ is in two spatial dimensions.
Clearly,
the wavefunction $\Phi$
is an eigenfunction of the many-particle angular-momentum
operator $\hat L_z$ with the total angular momentum $L=Nl$.

\begin{figure}[!]
\includegraphics[width=\columnwidth,angle=0]{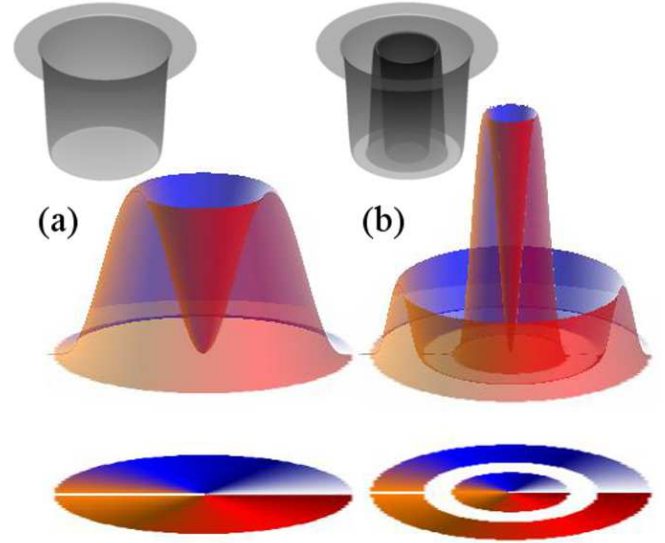}
\caption{(Color online) 
(a) The standard vortex in a two-dimensional circular trap potential.
The bosons carry the same angular momentum, $L/N=1$.
(b) Many-body vortex of the first kind in a circular trap split by a radial barrier.
Although the bosons carry the same angular momentum, $L/N=1$,
the wavefunction is fragmented, many-body in nature and composed of an inner and outer clouds.
See text for more details.
Shown are the densities, the bottom panels plot the phase.
The insets (in gray) depict the trapping potentials.}
\label{f1}
\end{figure}

In the present work we go beyond the current paradigm for vortices and
introduce vortices of the many-body kind, 
or many-body vortices (MBVs)
as they shall be referred to below.
MBVs are objects made of {\it spatially-partitioned} clouds 
and carry definite total angular momentum.
Two kinds of MBVs are discussed.
MBVs of the first kind are at the global minimum of the energy 
for states with definite total angular momentum
and where all the bosons carry the same angular momentum.
MBVs of the second kind are excited states
in which macroscopic fractions of bosons carry different angular momenta.
We devise a phase diagram for MBVs based on a simple model, 
and use it to predict and analyze the parameters of their occurrences. 
MBVs are fragmented rather than condensed objects,
and can only be described beyond mean-field theory.
Fragmentation is a many-body property of BECs derived 
from the eigenvalues of the reduced one-body 
density matrix \cite{Lowdin,Penrose,RDMbook}. 
If there is only one macroscopic eigenvalue the system is referred to as condensed,
whereas if there are two or more such eigenvalues the system is said to be fragmented.
Fragmentation of BECs has drawn much attention, see, e.g., 
\cite{nozieres:82,nozieres:96,Spekkens,Pit_Stri,BMF,ExBMF,2B,MCHB,Erich,fg1,fg2,fg3,fg4,fg5}. 

Consider a repulsive trapped BEC made of $N=100$ bosons in two spatial dimensions.
The many-boson Hamiltonian is given by
$\hat H(\br_1,\ldots,\br_N) = \sum_{j=1}^N [\hat T(\br_j) + V(\br_j)] + \sum_{j<k} W(\br_j-\br_k)$. 
$\hat T(\br)$ is the kinetic-energy operator, 
$V(\br)$ the trap potential,
and $W(\br-\br')$ the interparticle interaction.  
The short-range interaction between the bosons is modeled by a Gaussian 
function \cite{Gauss,222}
$W(\br-\br') = \lambda_0 \frac{e^{-(\br-\br')/2\sigma^2}}{2\pi\sigma^2}$
with a width $\sigma=0.25$. 
The interaction parameter is $\lambda_0=0.02$.
We employ dimensionless units,
which are readily obtained by dividing the Hamiltonian by $\frac{\hbar^2}{md^2}$,
where $m$ is the mass of a boson and $d$ is a length scale.
Realistic experimental parameters for the systems studied
below are given in \cite{units}.

We begin with the standard vortex in the potential of Fig.~\ref{f1}a,
computed with the Gross-Pitaevskii wavefunction (\ref{GP_Wf}) 
for angular momentum $L/N=1$.
Here $V(\br) = V_c(\br)$, where $V_c(\br) = \{ 200 e^{-(r-r_c)^4/2}, r \le r_c = 9; 200, r>r_c\}$
is a flat trap which has the shape of a ``crater'' of radius $r_c$.
We have chosen a flat potential $V_c(\br)$ 
in order to allow the BEC to fill in the full area.
This can readily be seen in Fig.~\ref{f1}a.

\begin{figure}[!]
\includegraphics[width=0.9\columnwidth]{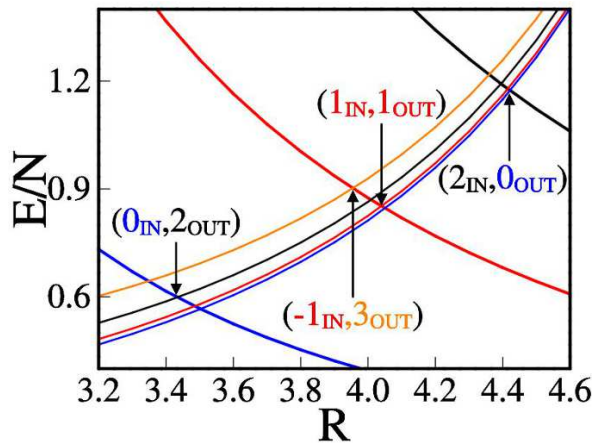}
\caption{(Color online) 
Ground-state phase diagram of many-body vortices in a circular trap.
A model based on the Gross-Pitaevski mean-field theory is employed.
The number of bosons is $N=100$ and the interaction parameter $\lambda_0=0.02$.
The energies of states in the inner (IN) disk and outer (OUT) annulus
parts of the trap are, respectively, decreasing and increasing as a function of the radius $R$.
The four (three) curves represent, from bottom to top,
states in the OUT (IN) part
with total angular momentum $L/N=0$ (solid blue), $1$ (solid red), $2$ (solid black),
and $L/N=3$ (solid orange).
States with angular momentum $\pm L/N$ are degenerate.
The crossings of the IN and OUT curves in the mean-field diagram 
marks the parameters where the MBVs occur.
For these radii, 
maximal fragmentation on the many-body level is encountered.
See text for more details.
All quantities are dimensionless.}
\label{f2}
\end{figure}

The central question we address in this work
is how to go beyond the standard mean-field vortex (\ref{GP_Wf})?
The answer, as we shall see below,
has to do with partitioning the BEC in space into two parts, 
thereby introducing many-body effects.
In as much as splitting a BEC into two, left--and--right clouds 
leads to many-body physics and even fragmentation, see, e.g., \cite{Spekkens},
we hereby propose to analogously do that with vortices. 
To this end, 
the BEC is now placed in the
radially-split trap shown in Fig.~\ref{f1}b.
The explicit form of the one-body potential is given by
$V(\br) = V_c(\br) + V_b(\br)$,
where $ V_b(\br) = 200 e^{-2(r-R)^4}$
is a ringed-shaped radial barrier of radius $R$. 
The BEC has now, in principle, two spatially-partitioned areas to occupy.
What would happen?

The split potential $V(\br)$
can be considered as made of two distinct parts,
an inner disk and an outer annulus,
separated by a radial barrier centered at $r=R$.
In the forthcoming analysis, 
this observation serves in forming a model.
Obviously, the energy of the system of total angular momentum $L/N=1$
changes with the barrier's radius $R$.
Furthermore, the location of the vortex in the split potential also changes with $R$.
To facilitate our discussion,
the energies of a vortex in the inner disk and outer annulus
as a function of $R$
are plotted in Fig.~\ref{f2}.
These energies are computed numerically 
using the ansatz (\ref{GP_Wf}),
i.e., by solving the Gross-Pitaevskii mean-field equation 
in the disk and, separately, in the annulus \cite{Note_No_Inter}.
For larger $R$, 
the system
is located in the inner disk, 
and looks much like the vortex shown in Fig.~\ref{f1}a.
For smaller $R$, 
it is energetically favorable for the system
to be located in the outer annulus.
Now, the system's lowest state of definite angular momentum $L/N=1$
is a vortex state in the outer annulus rather than in the inner disk.

The inner disk and outer annulus become energetically equivalent
at the barrier's radius $R=4.04$,
see the crossing point marked as $\mathrm{(1_{IN},1_{OUT})}$ in Fig.~\ref{f2}.
One might expect then 
the BEC to occupy the entire trap and,
consequently, 
for many-body effects to set in.
We are now in a need for a suitable theoretical framework
to cope with such many-body effects.

To go beyond the standard paradigm for vortices (\ref{GP_Wf}),
let us suppose that the system of many identical bosons
occupies not one but two, orthogonal one-particle functions, $\phi_1(\br)$ and $\phi_2(\br)$.
These functions carry the angular momenta $l_1$ and $l_2$, respectively.
The system's most general wavefunction takes on the form 
\beqn\label{MB_Wf}
& & \phi_1(\br) = e^{il_1\varphi} f_1(r), \quad 
\phi_2(\br) = e^{il_2\varphi} f_2(r), \quad \longrightarrow \nonumber \\
& & |\Psi\rangle = \sum_{n_1=0}^N C_{n_1} |n_1,n_2\rangle, \quad n_1+n_2=N. \
\eeqn
In (\ref{MB_Wf}),
the $C_{n_1}$ are expansion coefficients and
$|n_1,n_2\rangle$ are Fock states (permanents)
with $n_1$ bosons in the one-particle function $\phi_1(\br)$ and $n_2$ bosons in $\phi_2(\br)$.
For the wavefunction $|\Psi\rangle$
to be
an eigenfunction of the many-particle angular-momentum operator $\hat L_z$, 
let $l_1=l_2=l$.
In this case, 
there is in principle no restriction on the expansion coefficients $C_{n_1}$.
Namely, any distribution of the bosons
between the two one-particle functions $\phi_1(\br)$ and $\phi_2(\br)$
has the total angular momentum $L/N=l$.
Let us investigate now the BEC in the split trap of Fig.~\ref{f1}b
with the many-body ansatz (\ref{MB_Wf}).

To calculate the optimal many-body wavefunction (\ref{MB_Wf}) of the BEC,
namely to determine the one-particle functions $\phi_1(\br)$ and $\phi_2(\br)$
and expansion coefficients $C_{n_1}$,
we employ the multiconfigurational time-dependent Hartree 
for bosons (MCTDHB) method \cite{MCTDHB1,MCTDHB2,book2}. 
The MCTDHB method has been shown to produce accurate many-body solutions 
in various applications, see, e.g., 
\cite{BJJ,MCTDHB_OCT,Benchmarks,LC_NJP,MCTDHB_2D_3D,MCTDHB_2D_3D_dyn,2DBEC}. 
Explicitly,
we employ imaginary-time propagation within the 
the recursive MCTDHB (R-MCTDHB) \cite{R_MCTDHB}
and MCTDHB \cite{MCTDHB_Package} software packages, 
augmented by angular-momentum projection operators. 
For our investigations, 
a square box of size $[-12,12) \times [-12,12)$ and
spatial grid of size $128 \times 128$ are used and found to converge the results
to the accuracy given below.

As a first application, we have computed with the many-body ansaz (\ref{MB_Wf}) 
the standard vortex and found the mean-field description (\ref{GP_Wf}) to be valid 
in the flat potential trap of Fig.~\ref{f1}a.
Explicitly, the BEC is condensed to more than $99.9\%$.
We then turned to the split potential of Fig.~\ref{f1}b,
and to the crossing point at radius $R=4.04$ marked as $\mathrm{(1_{IN},1_{OUT})}$ in Fig.~\ref{f2}.
Indeed, as was expected,
the density is spread now in the disk {\it and} the annulus,
see Fig.~\ref{f1}b,
while the entire BEC has one and the same phase.
We remind that the total 
angular momentum is $L/N=1$.

The object shown in Fig.~\ref{f1}b has appealing properties and deserves a deeper examination.
First, its energetics.
Of all states carrying definite total angular momentum $L/N=1$,
it has the lowest energy in the trap.
Second, it is an `extended' object which combines two
spatially-partitioned parts.
The inner part looks much like 
the standard vortex,
the outer part as an annulus vortex state.
We stress that this is one and the same quantum object with a single many-body wavefunction!
Third, analysis of the many-body wavefunction reveals
that the object in Fig.~\ref{f1}b is actually fragmented rather than condensed,
with natural orbitals' occupations of $n_1=50.45\%$ and $n_2=49.55\%$.
Fourth, the two fragments are coupled, see for more details the Supplemental
Information \cite{SI}.
Summarizing all these properties,
we shall henceforth refer to this novel quantum object as 
an MBV of the first kind.

The idea of an MBV -- a spatially-partitioned and fragmented quantum object, carrying macroscopic definite
total angular momentum  --
can be further pursued.
This is because now a degree-of-freedom to distribute
the angular momentum between two parts is available.
Let us come back 
to the many-body wavefunction (\ref{MB_Wf}).
The second possibility to the requirement 
from $|\Psi\rangle$
to carry definite total angular momentum $L$
can be met for $l_1 \ne l_2$.
This dictates that only a single term in (\ref{MB_Wf}) can have
a non-vanishing expansion coefficient $C_{n_1}$, 
say the term with $n_1=n$.
In that case the total angular momentum of
$|\Psi\rangle$ is $L/N = (n/N) l_1 + (1-n/N) l_2$.
This is a general argument.
We shall now demonstrate how it manifests itself 
with a BEC in the split trap
of  Fig.~\ref{f1}b.

We return to the model of the split trap in terms of an
inner disk and outer annulus,
and compute the states of angular momentum $L/N=0,1,2,3$
using the Gross-Pitaevskii wavefunction (\ref{GP_Wf}).
Fig.~\ref{f2} plots the energies of these states
as a function of the radius $R$.
As the radius increases, the energies of states in the disk decrease,
and those of the states in the annulus increase.
States with higher angular momentum $L/N$ are higher in energy,
and those with $\pm L/N$ are degenerate.

There are many intersection points of the Gross-Pitaevskii energy curves for
the inner and outer states.
We concentrate on three such intersection points,
marked as $\mathrm{(0_{IN},2_{OUT})}$,
$\mathrm{(-1_{IN},3_{OUT})}$, and
$\mathrm{(2_{IN},0_{OUT})}$ in Fig.~\ref{f2}.
These occur at completely different radii, $R=3.43$, $3.955$, and $4.42$, respectively.
What is common to these three intersection points?
First, by definition, that the Gross-Pitaevskii energy of a BEC in the inner disk is equal
to that in the outer annulus.
Second, that the BEC in the inner part has {\it different} angular momentum
than the BEC in the outer part.
And third, that the total angular momentum of
a fragmented state equally distributed between such states
would be $L/N=1$.

\begin{figure*}[t]
\includegraphics[width=1.8\columnwidth,angle=0]{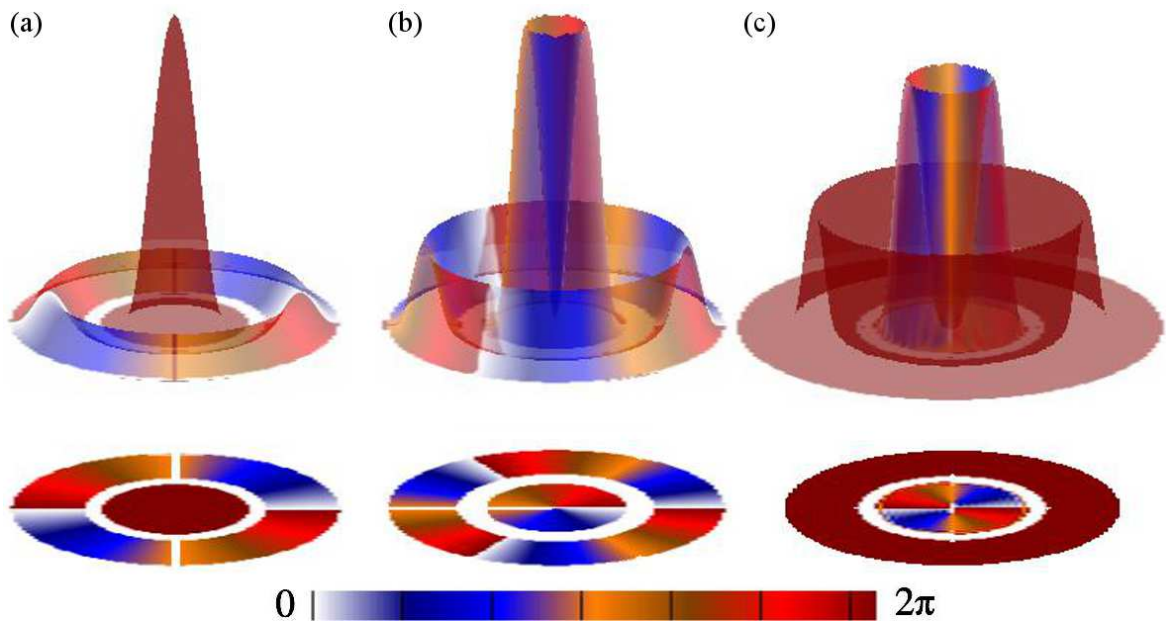}
\caption{(Color online) Many-body vortices of the second kind.
The total angular momentum is essentially $L/N=1$.
The angular momenta $l_1$ and $l_2$ of
the fragments in the disk and annulus parts are, respectively: 
(a) $l_{1}=0$ and $l_{2}=2$;
(b) $l_{1}=-1$ and $l_{2}=3$; and
(c) $l_{1}=2$ and $l_{2}=0$.
Shown are the densities, the bottom panels plot the phases.
Observe the different phases in the inner and outer parts.
See text for more details.}
\label{f3}
\end{figure*}

With the above preparatory analysis at hand,
we have employed the many-body ansatz (\ref{MB_Wf})
at the respective radii $R=3.43$, $3.955$, and $4.42$.
We have searched for the states lowest-in-energy when 
allowing the bosons to be distributed between two one-particle functions,
$\phi_1(\br)$ and $\phi_2(\br)$,
with angular momenta $l_{1}=0$ and $l_{2}=2$,
$l_{1}=-1$ and $l_{2}=3$, 
and $l_{1}=2$ and $l_{2}=0$, respectively.
In all three cases,
as expected from the conservation of angular momentum,
a single term of the many-body expansion (\ref{MB_Wf}) emerges.
Furthermore, in all three cases,
in line with the predictions of the phase diagram in Fig.~\ref{f2},
the term lowest-in-energy is the one with roughly $n_1=n_2=50\%$ natural orbitals' occupations
(the respective numerical values are: $n_2=48.0\%$, $n_1=52.0\%$ for the orbitals with $l_{1}=0$ and $l_{2}=2$; 
$n_1=50.0\%$, $n_2=50.0\%$ for the orbitals with $l_{1}=-1$ and $l_{2}=3$;
and $n_2=47.0\%$, $n_1=53.0\%$ for the orbitals with $l_{1}=2$ and $l_{2}=0$).
Fig.~\ref{f3} depicts the densities of these novel objects,
all spatially-partitioned, macroscopically fragmented, 
and carrying definite total angular momentum of essentially $L/N=1$.
Note the phases of each object which are distinct in the inner and outer parts.
Nonetheless, each of them is a quantum object with a single many-particle wavefunction,
see the Supplemental Information for more details \cite{SI}. 
We shall henceforth refer to these quantum objects as 
MBVs of the second kind.

A comparative discussion on the energetics and stability of MBVs is in place.
The MBVs of the first kind (Fig.~\ref{f1}b) are at the global minimum of energy;
They correspond to a fragmented ground state of a BEC 
with total definite angular momentum $L/N$
in the split trap of Fig.~\ref{f1}b. 
The MBVs of the second kind (Fig.~\ref{f3}) are macroscopically fragmented excited states \cite{ExBMF}.
Consider the MBV associated with the crossing point 
$\mathrm{(0_{IN},2_{OUT})}$ in the phase diagram of Fig.~\ref{f2}.
The global minimum of energy for the corresponding barrier's radius $R=3.43$
and angular momentum $L/N=1$
is the annulus vortex state
in the outer part of the trap. 
The wavefunctions of the MBV and annulus vortex state 
are, of course, orthogonal.
Furthermore, if trap imperfections are to introduce coupling
between the two states, 
we expect the tunneling of $N/2$ of the particles from the inner disk to the outer annulus, 
along with the exchange of angular momentum between the two parts,
to be a very slow process. 
This would render even in case of trap imperfections the MBV a long-lived quantum state.
The same analysis holds for the MBV associated
with the crossing point 
$\mathrm{(-1_{IN},3_{OUT})}$ at barrier's radius $3.955$.
Finally, for the MBV associated
with the crossing point 
$\mathrm{(2_{IN},0_{OUT})}$ at barrier's radius $4.42$,
the global minimum of energy 
for angular momentum $L/N=1$
is a vortex state
in the inner disk part of the trap. 
Here as well, 
in case of trap imperfections
we expect the tunneling of
$N/2$ particles from the outer annulus to the inner disk
to be a slow process,
rendering thereby
the MBV a long-lived quantum state.

In conclusion,
in the present work we go
beyond the standard paradigm of a vortex in Bose-Einstein condensates,
which is a localized object well described by mean-field theory,
and introduce vortices of the many-body kind. 
The MBVs are made of spatially-partitioned clouds,
carry definite total angular momentum, 
and are fragmented rather than condensed quantum objects, 
describable only beyond mean-field theory.
Two kinds of MBVs are discussed.
MBVs of the first kind are at the global minimum of the energy
for states of definite total angular momentum
and comprised of bosons carrying the same angular momentum.
MBVs of the second kind are fragmented excited states
in which macroscopic fractions of bosons carry different angular momenta.
A phase diagram based on the solutions of the Gross-Pitaevskii
equation in the inner and outer parts of the trap
is instrumental in predicting the parameters where MBVs occur. 
The most recent experiment \cite{Dalibard_exp} utilizing a
spatially-split circular trap like in Fig.~\ref{f1}
encourages us to anticipated that many-body vortices
will be further studied theoretically and
searched for experimentally.

\section*{Acknowledgements}

We thank Alexej Streltsov and Lorenz Cederbaum for discussions.
Partial financial support by
the DFG is acknowledged.
Computation time on the Cray XE6 system Hermit and
the NEC Nehalem cluster Laki at the HLRS, and the bwGRiD cluster
are gratefully acknowledged.

\onecolumngrid
\twocolumngrid



\onecolumngrid

\hglue 0.75 truecm

\section*{Supplemental Information}

We have introduced in the main text the spatially-partitioned many-body vortices.
Because the potential holding them has a high radial barrier,
the natural occupation numbers of the two fragments are nearly 50\% each.
An important issue in this case is the coupling between the two fragments.
An instrumental way to answer this question unequivocally,
is provided by investigating  
the so-called pathway from condensation to fragmentation which
the standard, mean-field vortex undergoes when
the radial potential barrier is ramped up all the way to the spatially-partitioned many-body vortex.
The two inter-connected questions are the subject of the Supplemental Information.

\section{Natural orbitals of spatially partitioned many-body vortices}

\addtocounter{figure}{-3}

\renewcommand{\figurename}{Figure S\hglue -0.115 truecm}

\begin{figure}[h]
\vglue -0.2 truecm
\includegraphics[width=0.85\columnwidth,angle=0]{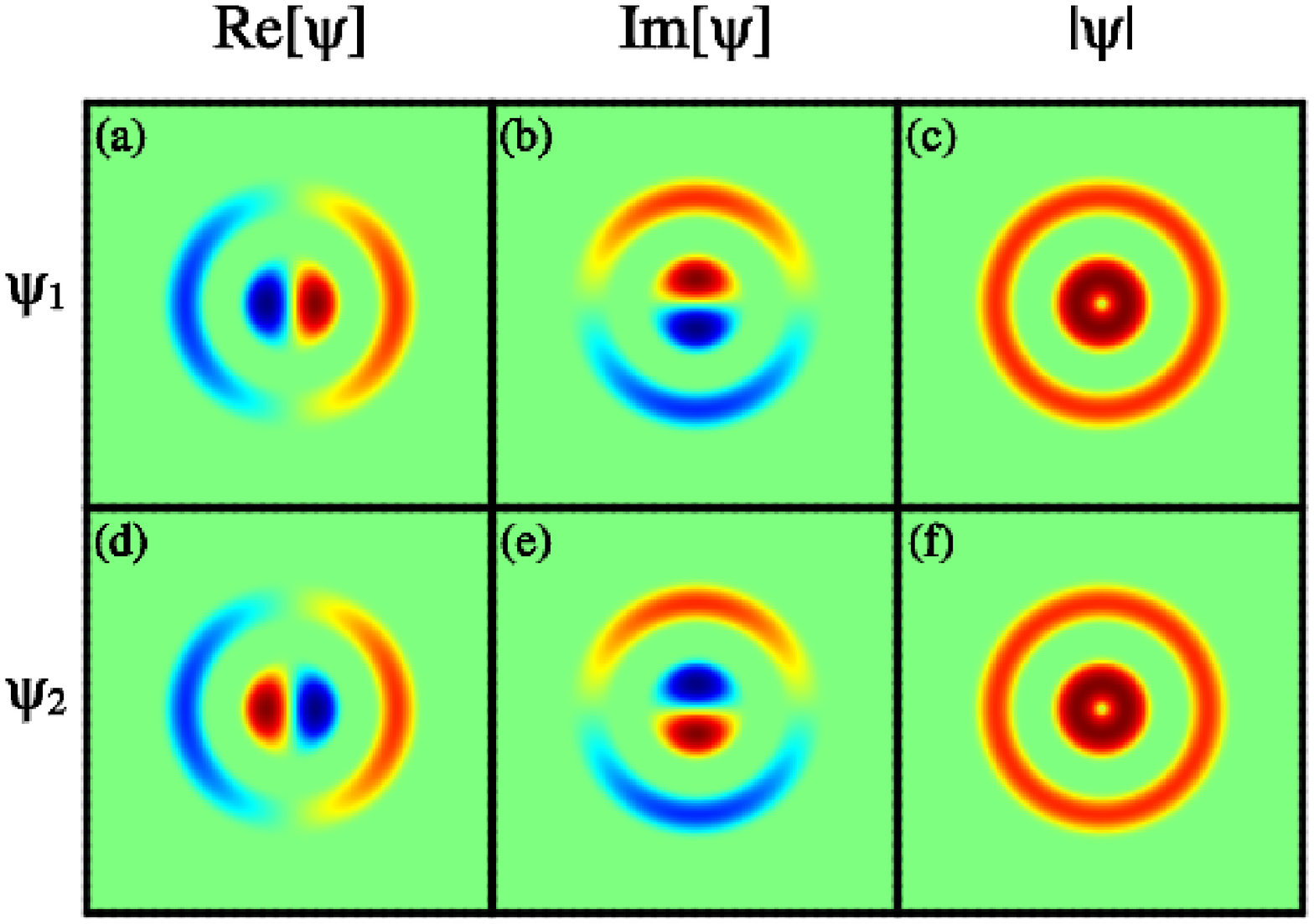}
\vglue -5.5 truecm
\caption{(Color online) 
The many-body vortex natural orbitals delocalize over the inner disk and outer annulus regions 
of the trap, which are coupled. 
Shown are the real, imaginary, and absolute value of the natural orbitals. 
The natural orbitals are seen to be continuous, complex-valued functions.
Each delocalized fragment carry the same angular momentum per particle, 
$l_1=l_2=1$.
All quantities are dimensionless.}
\label{sif1}
\end{figure}

\renewcommand{\figurename}{Figure S\hglue -0.10 truecm}

\begin{figure}[h]
\vglue -2.5 truecm
\includegraphics[width=0.85\columnwidth,angle=0]{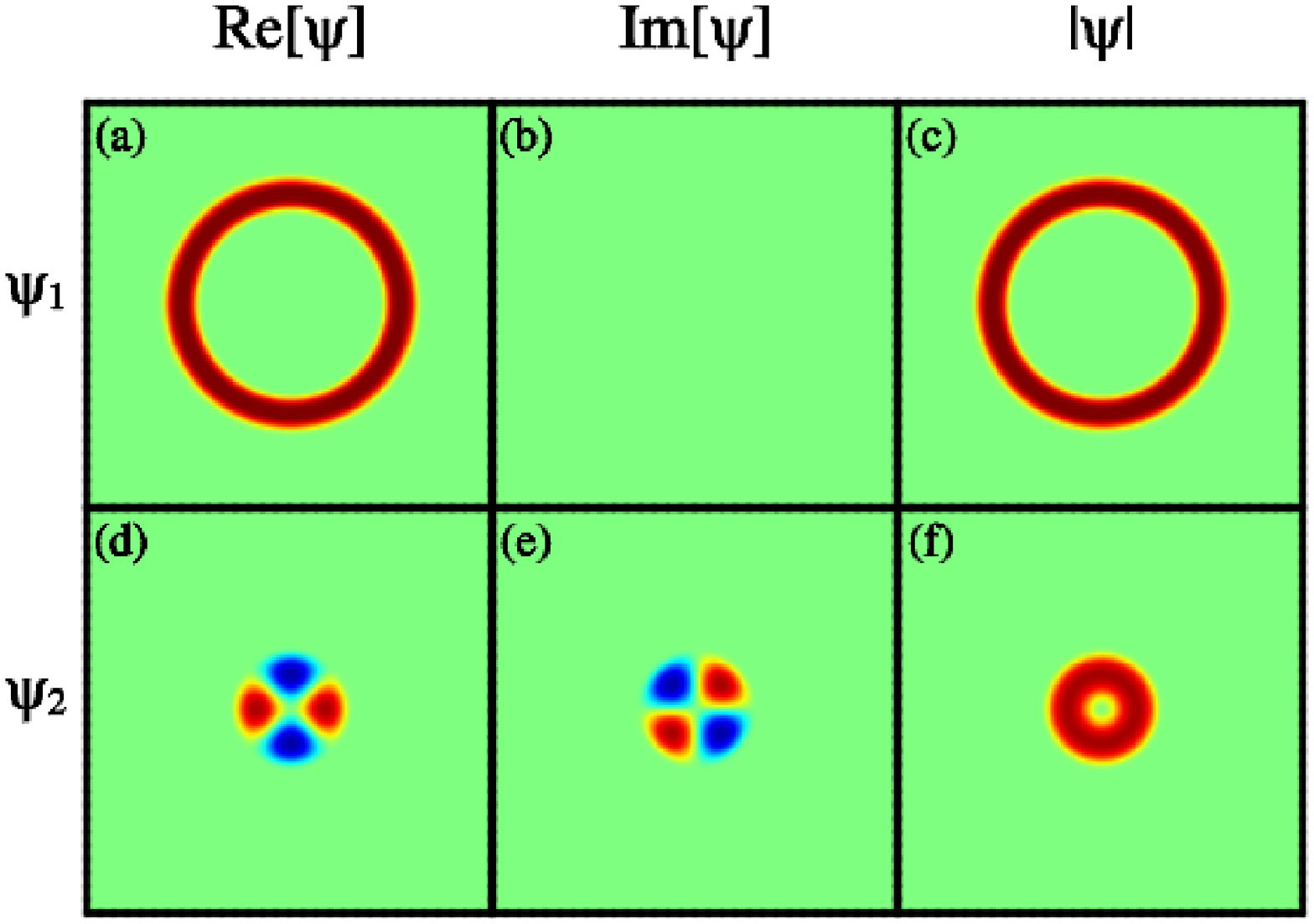}
\vglue -5.5 truecm
\includegraphics[width=0.85\columnwidth,angle=0]{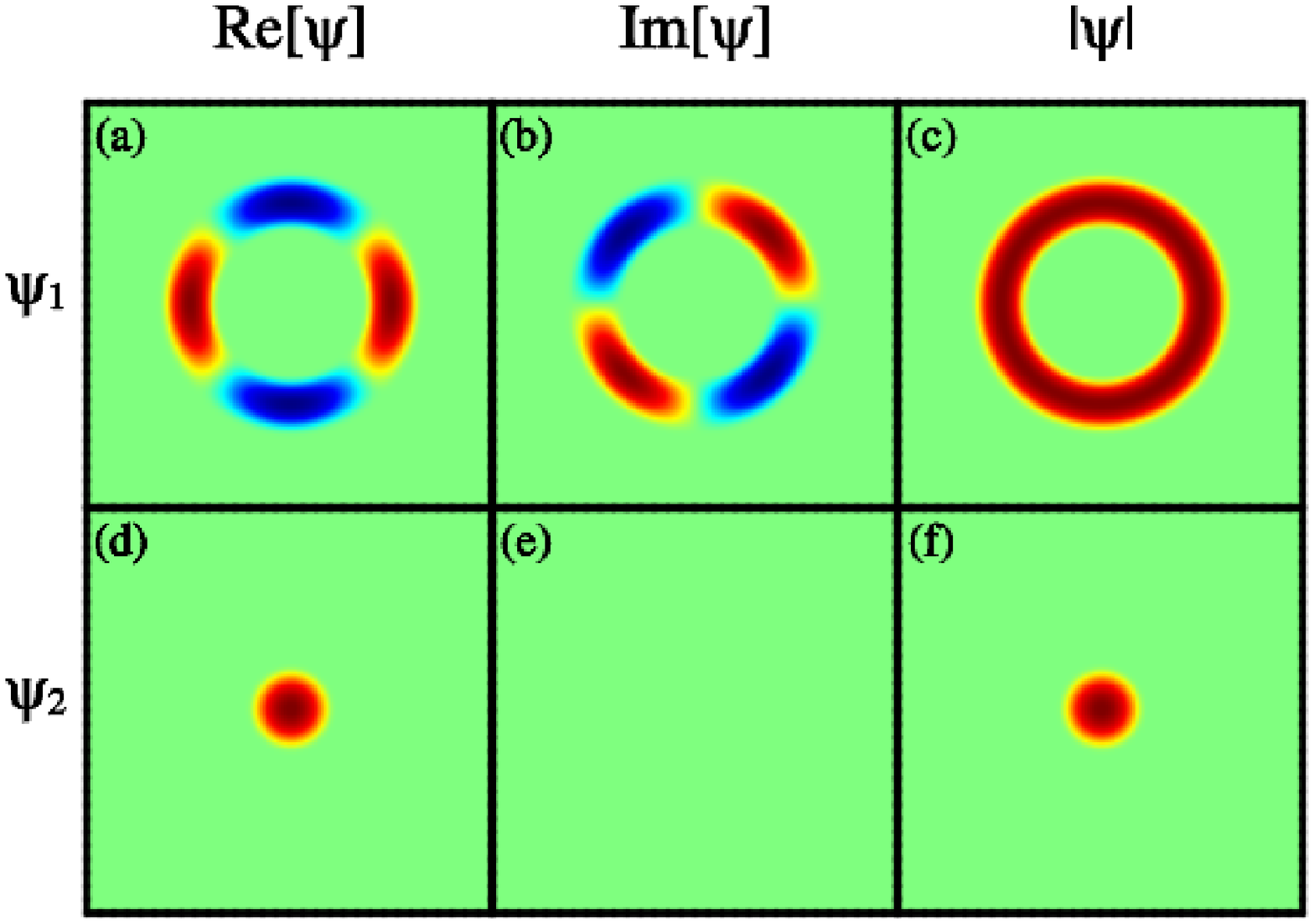}
\vglue -5.8 truecm
\caption{(Color online) 
The many-body vortex natural orbitals for 
different angular momenta (top: $l_1=2$, $l_2=0$; bottom: $l_1=0$, $l_2=2$) 
localize in the inner disk and outer annulus regions 
of the trap, which are nonetheless coupled for a finite barrier height.
Shown are the real, imaginary, and absolute value of the natural orbitals, 
which are seen to be continuous functions.
All quantities are dimensionless.}
\label{sif2}
\end{figure}

A few remarks are instructive.
First, we recall and stress that there is no general theorem requiring the natural orbitals
of a (repulsive) BEC to be localized.
As a consequence of the above, 
the shapes of the radial parts of the two fragments 
are {\it a priori} not constrained.
The two natural orbitals in the case $l_1=l_2=1$ (MBV of the first kind),
see Fig.~S1, are found 
by the many-body computation to be delocalized 
over the disk and annulus regions.
In the case $l_1 \ne l_2$ (MBV of the second kind),
the two natural orbitals are found 
by the many-body computation to be localized in the disk and annulus regions,
see Fig.~S2.\\

Second, since the many-body vortex is described by a many-body wavefunction
[see Eq.~(2) of the main text],
each of the fragments has its own phase, $e^{i l_1\varphi}$ and $e^{i l_2\varphi}$, respectively,
throughout all space.
This is not the standard Gross-Pitaevskii vortex,
with a single phase [see Eq.~(1) of the main text],
that has been exclusively explored in the literature.
Each of the fragments is a continuous function of the coordinates.
For $l=0$ this function is real valued, 
whereas for $l \ne 0$ it is complex valued.

Finally, the reason why the occupation numbers of the many-body vortices
can differ from exactly 50\% each is two fold.
Chiefly, because the barrier is not ``high enough''.
This is the topic of the subsequent section, see Fig.~S3 therein.
Furthermore, the ``optimal'' radius $R$ of the barrier 
for which the disk and annulus regions 
are energetically equivalent is determined according to a mean-field model, namely,
in a ``brute-force'' absence of any coupling between the inner and outer parts.
In realty, the two are coupled and the system is
solved on the many-body level. 

\newpage

\section{From the standard vortex to the spatially partitioned many-body vortex}

\renewcommand{\figurename}{Figure S\hglue -0.10 truecm}

\vglue -0.3 truecm
\begin{figure}[h]
\includegraphics[width=0.75\columnwidth,angle=0]{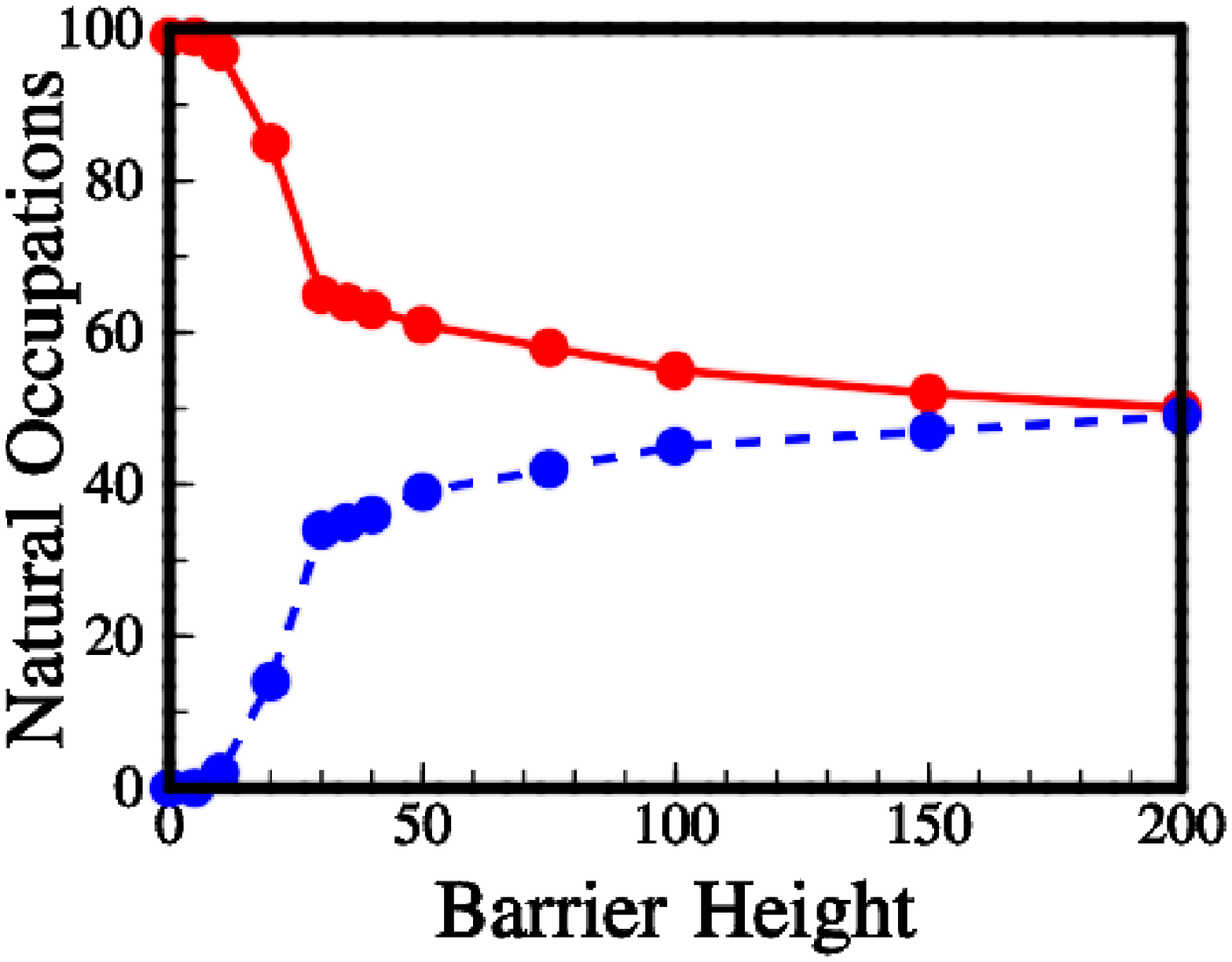}
\caption{(Color online) 
The \textit{ground state} of the BEC in the circular trap
as a function of the barrier height.
All bosons carry the same angular momentum, $l_1=l_2=1$.
Shown are the occupation numbers (in percents).
The ground state is no longer condensed even for low barrier heights.
The Gross-Pitaevskii theory ceases to be applicable and a many-body description of the state is necessary.
With increasing barrier's height,
the system gradually develops fragmentation.
Even for the highest barrier considered (the barrier height used in the main text, see Fig.~1),
we have a single many-body system,
and not two decoupled systems.
All quantities are dimensionless.}
\label{sif3}
\end{figure}


\begin{thebibliography}{99}
\bibitem{units} As a concrete example we choose $d=2 \mu$m 
and $^{87}$Rb atom as a boson.
We point out that other realistic choices can be made.
The unit of energy $\frac{\hbar^2}{md^2}$ corresponds to $29.1$Hz.
The relation between the dimensionless interaction parameter
$\lambda_0$, and the scattering length $a_s$
and the transverse confinement $l_z=\sqrt{\frac{\hbar}{m\omega_z}}$ along the $z$ axis 
is given by $\lambda_0 = 2 \sqrt{2\pi} \frac{a_s}{l_z}$,
where \cite{GSL} is used.
For $\omega_z = 2\pi \cdot 897$Hz (the aspect ratio is $\frac{l_z}{2r_c} = \frac{1}{100}$) 
and $a_s=1.44$nm, 
we get $\lambda_0=0.02$.
\end{thebibliography}
\end{document}